\documentclass[conference]{IEEEtran}
\IEEEoverridecommandlockouts
\usepackage{cite}
\usepackage{amsmath,amssymb,amsfonts}
\usepackage{algorithmic}
\usepackage{graphicx}
\usepackage{comment}
\usepackage{textcomp}
\usepackage{hyperref}
\usepackage{xcolor}
\def\BibTeX{{\rm B\kern-.05em{\sc i\kern-.025em b}\kern-.08em
    T\kern-.1667em\lower.7ex\hbox{E}\kern-.125emX}}
\begin{document}

\title{Metamorphic Testing for Smart Contract Validation: A Case Study of Ethereum-Based Crowdfunding Contracts}

\author{
\IEEEauthorblockN{1\textsuperscript{st} Irving Jared Villanueva}
\IEEEauthorblockA{
\textit{Computer Science Department} \\
\textit{East Carolina University}\\
Greenville, USA}
\and
\IEEEauthorblockN{2\textsuperscript{nd} Madhusudan Srinivasan\textsuperscript{*}}
\IEEEauthorblockA{
\textit{Computer Science Department} \\
\textit{East Carolina University}\\
Greenville, USA \\
Email: srinivasanm23@ecu.edu
\thanks{*Corresponding author}
}
\and
\IEEEauthorblockN{3\textsuperscript{rd} Faqeer Ur Rehman}
\IEEEauthorblockA{
\textit{Independent Researcher}\\
USA}
}




\maketitle

\begin{abstract}
Blockchain smart contracts play a crucial role in automating and securing agreements in diverse domains such as finance, healthcare, and supply chains. Despite their critical applications, testing these contracts often receives less attention than their development, leaving significant risks due to the immutability of smart contracts post-deployment. A key challenge in the testing of smart contracts is the oracle problem, where the exact expected outcomes are not well defined, complicating systematic testing efforts.

Metamorphic Testing (MT) addresses the oracle problem by using Metamorphic Relations (MRs) to validate smart contracts. MRs define how output should change relative to specific input modifications, determining whether the tests pass or fail. In this work, we apply MT to test an Ethereum-based crowdfunding smart contract, focusing on core functionalities such as state transitions and donation tracking. 

We identify a set of MRs tailored for smart contract testing and generate test cases for these MRs. To assess the effectiveness of this approach, we use the Vertigo mutation testing tool to create faulty versions of the smart contract. The experimental results show that our Metamorphic Relations (MRs) detected 25.65\% of the total mutants generated, with the most effective MRs achieving a mutant-killing rate of 89\%. These results highlight the utility of MT to ensure the reliability and quality of blockchain-based smart contracts.
\end{abstract}


\section{Introduction \& Motivation}
A smart contract on blockchain is a self-executing agreement where the terms between buyer and seller are written into code, enabling automatic execution without intermediaries. Running on decentralized networks such as Ethereum, smart contracts offer significant real-world benefits across sectors. In finance, smart contracts streamline loan issuance and automate escrow services~\cite{pal2021blockchain}. In addition, in supply chains, they improve transparency and reduce fraud through real-time tracking; in healthcare, they securely facilitate patient data sharing; and in legal contexts, they enforce agreements without lengthy dispute resolution processes~\cite{wang2019smart}~\cite{priyadarshini2022medchain}~\cite{mckinney2017smart}. However, thorough testing of smart contracts is essential because they are immutable once deployed and any errors or vulnerabilities can result in severe financial and security risks.

Testing smart contracts also presents unique challenges, particularly the test oracle problem, which arises from difficulties in defining a clear "expected outcome" against which to verify the actual output of the contract. Smart contracts operate in a distributed environment where execution depends on blockchain state, including block timestamps, gas prices, and network conditions, making it challenging to establish precise test oracles. Complex state transitions across multiple transactions can trigger cascading effects through interconnected contracts, complicating the prediction and verification of complete state changes. For example, in DeFi applications, a single trade affects multiple liquidity pools and token balances.

Smart contracts implement intricate business logic involving temporal, financial and access control properties, each requiring specific verification criteria. The immutable nature of deployed contracts further amplifies the importance of comprehensive testing. Moreover, external interactions through oracles and cross-contract calls introduce non-deterministic behavior and dependencies on external data sources, making it challenging to establish reliable test oracles for verifying the correctness of contract execution. This is especially complicated when contracts rely on external data sources or oracles, making expected results less predictable.

MT offers a solution by focusing on relationships, or Metamorphic Relations, between inputs and outputs rather than exact outcomes. For smart contracts, MT leverages these relationships to evaluate if the contract behaves consistently under different conditions, helping to identify faults even without a precise oracle. This approach provides a robust alternative to traditional testing methods, ensuring that smart contracts perform reliably and securely in various scenarios.

The main contributions of this study are as follows:

\begin{itemize}
    \item Designed 17 Metamorphic Relations (MRs) specifically for testing Ethereum-based crowdfunding smart contracts, targeting critical functionalities such as state transitions and donation tracking.
    
    \item Demonstrated the applicability of MT as a solution to the test oracle problem in smart contract testing.
    
    \item Introduced a structured approach for generating source and follow-up test cases to validate smart contract behaviors.
    
    \item Conducted extensive mutation testing using the Vertigo tool to evaluate the fault-detection capabilities of the proposed MRs.
    
    \item Showed that certain MRs achieved a high mutant-killing rate, with the most effective ones reaching up to 89\%.
    
    \item Highlighted the utility of MT in ensuring the reliability and correctness of key operations in smart contracts, such as lifecycle state transitions and consistent handling of donations.
    
    \end{itemize}
    
\section{Background}
In software testing, a test oracle is a mechanism or principle that determines whether the outcomes of a test are correct~\cite{chenmetamorphic}. It serves as a baseline to compare actual results against expected outcomes. However, in contexts such as smart contracts, defining precise expected outcomes can be challenging due to the complex state dependencies and external interactions involved. This issue, known as the test oracle problem, complicates systematic testing efforts. 

MT provides a solution to alleviate the test oracle problem. MT is a software testing approach designed to address situations where there is no clear expected outcome to compare the actual output against the known as the test oracle problem~\cite{segura2018metamorphic}. In such cases, MT verifies the consistency and correctness of the system by defining MRs which are predictable relationships between different sets of inputs and outputs. These MRs describe how an output should change in response to specific changes in input. If the system fails to satisfy these relationships, a fault is likely present.
For example, consider the function \( f(x) = x^2 \). Without knowing the exact output for \( f(3) \), we define a Metamorphic Relation: if \( x \) doubles, the output should quadruple. Testing this, we calculate \( f(3) = 9 \) and \( f(6) = 36 \). Since \( f(6) = 4 \times f(3) \), the function passes this consistency check, verifying its correctness indirectly.

MT has been applied across diverse domains to address challenges posed by the oracle problem. In machine learning, MT validates the robustness of models by examining consistent behavior under perturbations~\cite{xie2011testing}~\cite{nasr2024study}. In scientific computing, it ensures the accuracy of complex simulations, such as weather models or physical simulations, by checking for predictable relationships~\cite{lin2018exploratory}. MT is also widely used in financial systems to validate transaction consistency under varying conditions and in healthcare applications to test algorithmic outputs for medical diagnostics~\cite{
jafari2022evaluation}~\cite{ma2022metamorphic}. By leveraging MRs tailored to domain-specific requirements, MT ensures that systems operate reliably, even in scenarios lacking definitive expected outcomes. Expanding its use in blockchain and smart contract validation further demonstrates its adaptability and efficacy in guaranteeing system dependability.

A smart contract is a self-executing agreement in which the terms are encoded directly into the contract, running on blockchain networks like Ethereum. These contracts automatically fulfill specific conditions without intermediaries, bringing efficiency, transparency, and reduced transaction costs to various applications~\cite{zou2019smart}. In supply chains, for example, grocery chains such as Walmart use blockchain-based smart contracts to improve transparency, efficiency, and traceability. When fresh produce is packaged, a smart contract records key information such as harvest date, location, and batch details on the blockchain. As the product moves through each stage from the farm to the packaging, distribution centers, and covering factors such as storage conditions and transit times. This enables Walmart to trace the exact origin of any item within minutes, enhancing quality control and allowing for rapid response to food safety concerns. Using smart contracts, Walmart secures accurate and tamper-proof records throughout its supply chain, promoting transparency for consumers and regulatory bodies while reducing risks and operational inefficiencies~\cite{tan2018impact}.

\section{Related Works}

Yuan et al.~\cite{ji2024mt4sc} introduce a framework that utilizes user behavior sequences to effectively test smart contracts, particularly addressing the test oracle problem, which involves verifying contract correctness without predefined expected results. By simulating real-world user interactions through sequences of behaviors and constructing MRs, MT4SC aims to uncover contract faults that emerge from complex, transaction-driven state changes. The methodology involves defining user behavior sequences as dynamic test inputs, generating source and follow-up test cases, and using a simulation environment (Truffle and Ganache) to test contracts with state resets between executions. The MT4SC framework was tested on eight open source smart contract applications in various domains, focusing on fault detection effectiveness, code coverage, and time cost. The experimental results showed that MT4SC outperformed the baseline tools, achieving high fault detection rates, improved branch and statement coverage, and reasonable time costs. However, the approach involves designing effective MRs based on user behaviors and is complex and requires an in-depth understanding of contract functionality, which increases testing time and computational resource needs. Furthermore, detailed transaction sequencing limits scalability, making MT4SC resource intensive and potentially challenging to apply to more extensive and complex contracts. In contrast, our work specifically applies MT to an Ethereum-based crowdfunding smart contract and we utilizes defined MRs to systematically generate source and follow-up test cases to test core functionalities.

Li et el.~\cite{li2024characterizing} investigates Ethereum upgradable smart contracts (USCs) and their security implications. The methodology involves developing a tool called USCDetector, which identifies USC patterns by analyzing bytecode and transaction information, without relying on source code. USCDetector can detect multiple USC patterns, such as proxy and data separation, enabling large-scale analysis on the Ethereum blockchain. In experiments on more than 60 million contracts, USCDetector identified 10,218 USC upgrade chains and disclosed multiple security issues.  The results demonstrated high accuracy (96.26\%) in USC identification and significant security concerns among real-world contracts. However, the approach includes reliance on certain upgrade function keywords, limiting detection of non-standard USC implementations, and a computationally intensive process due to comprehensive data collection. In contrast, our work focuses on testing the core functionalities of a single smart contract using MT rather than analyzing large-scale patterns. Unlike USCDetector's emphasis on large-scale detection and security implications, our approach offers a more targeted, functionality-driven methodology for validating individual smart contracts, ensuring their correctness and reliability.

Jinlei et al.~\cite{sun2021mutation} explores mutation testing for integer overflow vulnerabilities in Ethereum smart contracts (ESC). The methodology involves designing five specialized mutation operators that target three main types of integer overflow: arithmetic, truncation, and signed overflow. These operators generate mutants that simulate realistic integer overflow vulnerabilities, allowing the authors to evaluate the effectiveness of integer overflow testing tools. An empirical study was conducted on 40 ESCs, generating 2,099 mutants, where 95\% were valid (compilable) and contained integer overflow vulnerabilities. The results demonstrated that the proposed operators successfully reproduced all 179 original integer overflow vulnerabilities, showing high effectiveness in detecting vulnerabilities that traditional testing approaches might overlook. However, the approach includes a high computational cost and limitations in detecting nonstandard overflow patterns, which may hinder scalability and practical application in larger-scale contract testing scenarios. In contrast, our work employs MT to validate an Ethereum-based crowdfunding smart contract, focusing on broader functional correctness rather than specific vulnerability types. Using tailored MRs and mutation testing, our approach systematically generates test cases to detect faults across various functionalities.

Ji et al.~\cite{ji2022test} propose an improved genetic algorithm (Iga-Sc) for the generation of test cases to achieve high coverage in the data flow testing of smart contracts. This approach integrates particle swarm optimization (PSO) principles to reduce the impact of randomness on genetic operations, thus enhancing the search for global optima. The methodology includes constructing a control flow graph (CFG) for smart contracts, performing data flow analysis to define variables and definition-use pairs, and applying Iga-Sc for efficient test case generation. This setup aims to provide high coverage of data flows by ensuring that variables and conditional paths are thoroughly tested in Solidity-based smart contracts. In the experimental results, Iga-Sc outperformed three baseline models ADF-GA, GA-C, and random testing (RT) in a data set of 30 smart contracts, achieving an average coverage of 89 2\% in pairs of definition-use. The experiment demonstrated that Iga-Sc reduced the number of iterations needed and execution time compared to the other models, confirming its higher efficiency and effectiveness. Specifically, Iga-Sc achieved a coverage improvement of 13.81\% to 32.54\% over the other models, with significantly fewer iterations and a shorter execution time. The study relies on a relatively small dataset of smart contracts with limited complexity, which may not fully represent the diversity of real-world contracts.  In contrast, our work applies MT to an Ethereum-based crowdfunding smart contract, focusing on detecting faults in core functionalities such as state transitions and donation tracking.

Barboni et al.~\cite{barboni2022sumo} presented a tool SuMo, a mutation testing tool specifically designed for Solidity smart contracts. It aims to improve the reliability of these critical pieces of code by assessing the effectiveness of existing test suites. Mutation testing works by introducing small, deliberate changes (mutations) to the original code. These mutations simulate potential programming errors. By observing how effectively the existing tests detect these mutated versions (mutants), developers can gain insights into the quality and coverage of their tests.

SuMo implements a diverse set of 44 mutation operators, including both traditional operators (e.g., arithmetic operator changes, boolean operator negation) and Solidity specific operators that target unique features of the language (e.g., modifiers, visibility modifiers, state mutability). These operators are carefully designed to cover a wide range of potential faults that can occur in smart contracts. The experimental results demonstrate the effectiveness of SuMo in identifying weaknesses in existing test suites for real-world Solidity projects. By applying SuMo to various open-source smart contracts, the researchers were able to uncover numerous mutants that were not detected by the original tests. This highlights the limitations of the existing test suites and provides valuable feedback to developers for improving their testing strategies. However in contrast, our work uses mutation testing tool to validate the effectiveness of MRs.

li et al.~\cite{li2019musc} designed MuSC, a tool designed for mutation testing of Ethereum Smart Contracts (ESCs). Mutation testing aims to evaluate the effectiveness of existing test suites by introducing small, deliberate changes (mutations) to the original code. These mutations simulate potential programming errors. By observing how effectively the tests detect these mutated versions (mutants), developers can assess the quality and coverage of their tests.

MuSC incorporates a set of mutation operators specifically tailored for the unique characteristics of Solidity, the primary language used for developing ESCs. These operators target various aspects of Solidity code, including arithmetic operations, boolean expressions, control flow statements, and data types. 

The experimental results demonstrate that MuSC can effectively generate a diverse set of mutants for ESCs. By applying MuSC to several real-world smart contracts, the researchers identified numerous mutants that were not detected by the original test suites. This highlights potential weaknesses in the existing testing strategies and provides valuable feedback to developers for improving the robustness of their smart contracts. In contrast, our work focuses specifically on testing smart contract using metamorphic testing and validate the effectiveness of MR using mutation testing.


\section{Methodology}
\label{sec:methodology}
In this section, we describe our methodology to test the Ethereum crowd-funding smart contract program using MT. Figure~\ref{fig:mt_process} shows the sequence of steps required to
perform MT. The following is the sequence of steps involved.
\begin{enumerate}
    \item Identify a set of MRs for testing the PUT. The MRs that we developed are described in section~\ref{label:MR Section}.
    
    \item Develop source test cases and follow-up test cases for each of the MRs. In Section~\ref{label:MR Section} we describe the source and follow-up test case.
    
\item The source and follow-up test cases for each MR are executed in the program. The results of the source and follow-up test cases are verified to identify if the corresponding MR is violated. The violation of MR indicates the presence of a fault in the smart contract program.

 \begin{figure}[ht]
\centering
\includegraphics[width=0.3\textwidth]{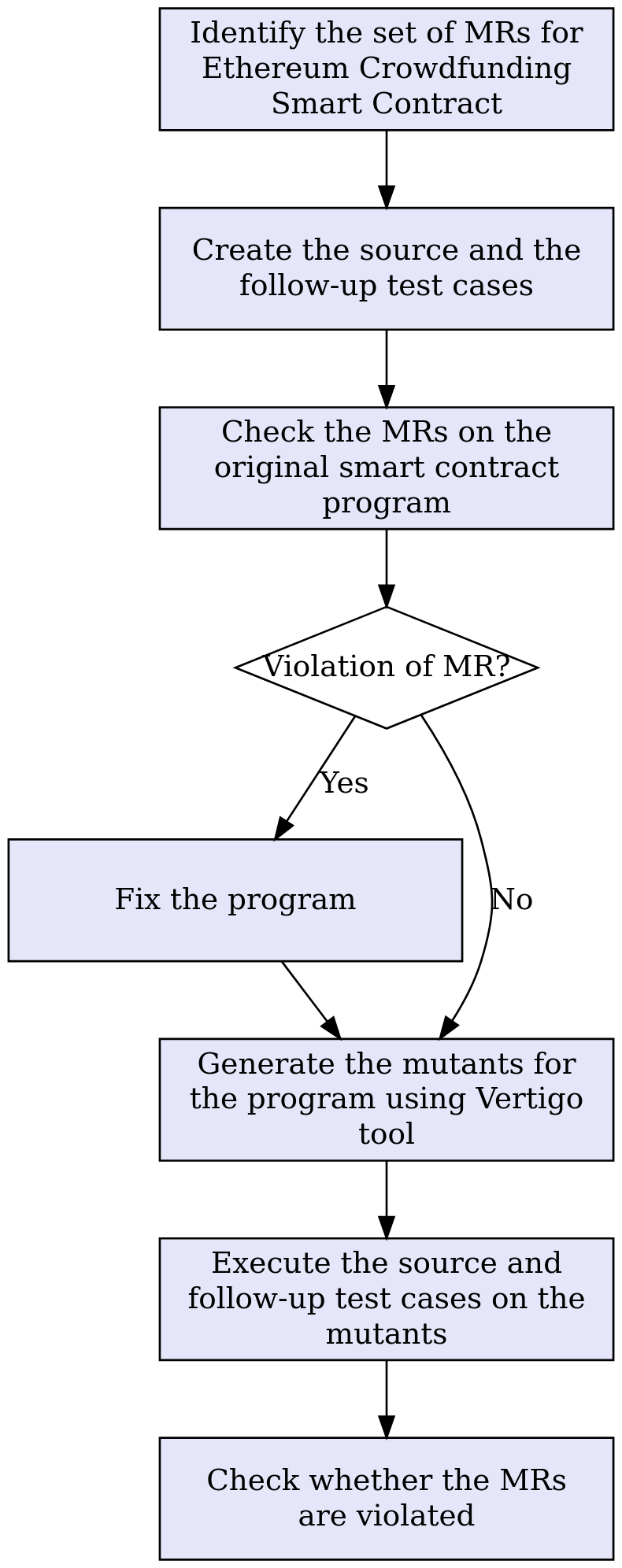}
\caption{MT Process for testing Ethereum Crowdfunding Smart Contracts}
\label{fig:mt_process}
\end{figure}
\end{enumerate}

\section{System Under Test}
In this work, SUT is an Ethereum-based crowdfunding application, represented by the CrowdfundingCampaign smart contract \footnote{\url{https://github.com/giobart/EtherCrowdfunding}}. This contract enables a structured and transparent crowdfunding process in which organizers can initiate a campaign, collect donations, and distribute funds to beneficiaries. The contract follows three main states: Started, Donation, and Ended. Initially, in the Started state, organizers must make an initial donation to activate the campaign. Once all organizers have donated, the contract transitions to the Donation state, allowing external users to contribute. Finally, after the campaign duration expires or all milestones are reached, the contract moves to the Ended state, permitting beneficiaries to withdraw their allocated funds.

The contract includes a milestone system managed by an auxiliary contract, CrowdfundingCampaignMilestoneSystem, which allows organizers to set milestones, reward contributions, and issue refunds for unachieved milestones after the campaign ends. Key functionalities of the SUT include donation options, where donors can contribute either uniformly across beneficiaries or specify amounts for each organizer's role. Also, it involves campaign initiation, milestone setup and donations that control the campaign's transition to active donation state, withdrawal management for beneficiaries, ensuring authorized withdrawals after the campaign ends, and Milestone Rewards, where organizers can set milestones to encourage donations or distribute rewards. Once all beneficiaries have withdrawn, organizers can terminate the contract, distributing any residual funds.

The contract uses Solidity libraries such as IterableAddressMapping for efficient address management, SafeMath for secure arithmetic, and AscendingOrderedStack to track donor rewards in ascending order. This structure ensures secure funding management, proper management of campaign states, and transparent donation tracking, making it a comprehensive system for testing crowdfunding functionalities on the Ethereum blockchain.

\section{Metamorphic Relations}
\label{label:MR Section}
\begin{itemize}
\item \textbf{MR1: State Transition Consistency}
Consider a source test case where the contract is initialized to a state \( S \), which is "STARTED," and no donations or state-altering actions have been performed. The state of the contract, \( T(S) \), is retrieved in the source test case to confirm that it remains in the "STARTED" state (\( T(S) = \text{STARTED} \)). This establishes the baseline behavior of the contract without any external interactions.

In the follow-up test case, starting from the initial state \( S \), valid operations are performed to transition the contract through its defined lifecycle. First, donations are made by the organizers, leading to a state transition from "STARTED" to "DONATION" (\( S \to S' \), where \( T(S') = \text{DONATION} \)). Finally, after the campaign duration elapses and the end campaign function is invoked, the state transitions from "DONATION" to "ENDED" (\( S' \to S'' \), where \( T(S'') = \text{ENDED} \)). These operations ensure that the contract progresses through the expected states: \( \text{STARTED} \to \text{DONATION} \to \text{ENDED} \), verifying consistent and accurate state transitions as defined in the contract's lifecycle.

\item \textbf{MR2: Donations Consistency}
Consider a source case where a single large donation \( D \) is made to the contract, and let \( A(D) \) represent the total amount raised by the contract after this donation. Create a follow-up scenario where the large donation \( D \) is split into multiple smaller donations \( D_1, D_2, \dots, D_n \) such that the sum of these smaller donations equals \( D \) (i.e., \( D = \sum_{i=1}^n D_i \)). Let \( A(D') \) represent the total amount raised by the contract after receiving the multiple smaller donations. We expect that \( A(D) = A(D') \), ensuring that the total amount raised is consistent regardless of how the donation amount is contributed.

\item \textbf{MR3: Deployment Consistency with Duplicate Beneficiaries}
Consider a source case where the contract is deployed with a unique set of beneficiaries, and let \( S(C) \) represent the initial state of the campaign after deployment. Create a follow-up case where the contract is deployed with duplicate beneficiaries, resulting in a modified deployment. Let \( S(C') \) represent the state of the campaign in this follow-up deployment. We expect that \( S(C) \neq S(C') \), indicating that the state of the campaign differs when duplicate beneficiaries are included, thus validating the contract's response to beneficiary duplication upon deployment.

\item \textbf{MR4: Organizer Donations Consistency}  
Consider a source test case \( S \) where the campaign is in the initial "STARTED" state with no donations made by organizers, and let \( T(S) \) represent the current state of the campaign. Create a follow-up test case \( S' \) where each organizer sequentially makes their initial donations. Let \( T(S') \) represent the campaign's state after these donations. We expect that \( T(S) \) should transition to the "DONATION" state only once all organizers have completed their initial donations, validating the requirement for organizer participation before progressing to the donation phase.

\item \textbf{MR5: Milestone Reward Consistency}  
Consider a source test case \( S \) where the contract balance \( B(S) \) is observed before any milestone is achieved. Create a follow-up test case \( S' \) where actions are executed to reach a predefined milestone. Let \( B(S') \) represent the contract balance after reaching the milestone. We expect that \( B(S') = B(S) - R \), where \( R \) is the milestone reward amount, ensuring that the contract balance decreases accurately by the reward amount upon milestone achievement.

\item \textbf{MR6: Consistency for Campaign Duration}  

Consider a source test case \( S \) where donations are made within the active campaign period, and let \( D(S) \) represent the successful acceptance of these donations. Create a follow-up test case \( S' \) where donation attempts are made both before the campaign start date and after the campaign end date. Let \( D(S') \) represent the outcome of these donation attempts outside the campaign period. We expect that \(D(S') \neq D(S)\),
where \( D(S') \) should reject donations made outside the campaign duration, ensuring the contract only accepts donations during the designated campaign timeframe. 


\item \textbf{MR7: Refund Consistency for Unreached Milestones}
Consider a source test case \( S \) where multiple milestones are set up within the campaign, with certain milestones intentionally set to be unreachable. Let \( R(S) \) represent the initial state where no refunds have been issued. Create a follow-up test case \( S' \) where refund requests are triggered after the campaign has ended for those milestones that were not achieved. Let \( R(S') \) represent the state after processing these refund requests. We expect that \( R(S) = R(S') \) where \( R(S') \) should reflect refunds issued to organizers for all unreached milestones, verifying that refunds are processed correctly post-campaign for unattained milestones.

\item \textbf{MR8: Consistency for Maximum Organizers or Beneficiaries Limit} 
Consider a source test case \( S \) where the contract is initialized with the maximum allowed number of organizers or beneficiaries, and let \( I(S) \) represent the successful initialization state of the contract. Create a follow-up test case \( S' \) where the contract initialization is attempted with a number of organizers or beneficiaries exceeding this maximum limit. Let \( I(S') \) represent the result of this initialization attempt. We expect that \( I(S) \neq I(S') \), where \( I(S') \) should fail, ensuring that the contract enforces the maximum limit on the number of organizers or beneficiaries, thus preventing initialization with an excess number.

\item \textbf{MR9: Consistency for Minimum and Zero Donations} 
Consider a source test case \( S \) where a donation is made that matches the minimum allowed amount, and let \( D(S) \) represent the success status of this donation. Create a follow-up test case \( S' \) where donation attempts are made with amounts below the minimum allowed value, including a zero-value donation. Let \( D(S') \) represent the outcome of these donation attempts. We expect that that \( D(S) \neq D(S') \), where \( D(S') \) should fail, ensuring the contract enforces the minimum donation requirement by rejecting any donations below the specified threshold, including zero-value contributions.

\item \textbf{MR10: Consistency Under Rapid State Transitions}  
Consider a source test case \( S \) where standard operations are performed sequentially, leading to state transitions at a regular pace, and let \( T(S) \) represent the stable progression of states in the contract. Create a follow-up test case \( S' \) where actions that trigger state changes such as donations or other relevant operations are executed in rapid succession. Let \( T(S') \) represent the resulting state progression under these accelerated conditions. We expect that \( T(S) = T(S') \), where \( T(S') \) should maintain consistency, with no errors or unexpected behavior, ensuring the contract handles rapid state transitions reliably and without introducing inconsistencies.

\item \textbf{MR11: Consistency Under Simultaneous Withdrawals}
Consider a source test case \( S \) where beneficiaries withdraw funds one at a time, and let \( W(S) \) represent the total funds withdrawn sequentially, ensuring proper tracking. Create a follow-up test case \( S' \) where all beneficiaries attempt to withdraw their funds simultaneously. Let \( W(S') \) represent the total funds withdrawn under these concurrent requests. We expect that \( W(S') = W(S) \), ensuring the contract correctly manages concurrent withdrawals by maintaining a consistent and accurate total withdrawal amount, despite simultaneous requests.

\item \textbf{MR12: Consistency for Last-Minute Donations}
Consider a source test case \( S \) where donations are made well within the active campaign duration, and let \( D(S) \) represent the successful acceptance of these donations. Create a follow-up test case \( S' \) where donation attempts are timed to coincide with the exact closing moments of the campaign. Let \( D(S') \) represent the contract's response to these last-minute donations. We expect that that \( D(S) = D(S') \) where \( D(S') \) will accept donations if they are made within the campaign duration and reject them if they occur after the campaign ends, ensuring the contract accurately handles donations near the campaign’s end time.

\item \textbf{MR13: Consistency for Overflow and Underflow Protection} 
Consider a source test case \( S \) where transactions are executed with typical, moderate donation amounts, and let \( A(S) \) represent the accurate accounting of these transactions within the contract. Create a follow-up test case \( S' \) where transactions involve extremely high or low values, pushing the boundaries to test for potential arithmetic overflows or underflows. Let \( A(S') \) represent the contract’s accounting response under these extreme conditions. We expect that \( A(S') = A(S) \), with the contract preventing any overflow or underflow errors, thereby ensuring accurate and consistent accounting across all transaction values.

\item \textbf{MR14: Consistency for Invalid Milestone Rewards} 
Consider a source test case \( S \) where milestones are set with typical, valid reward values, and let \( M(S) \) represent the successful establishment of these milestones. Create a follow-up test case \( S' \) where milestones are set with invalid reward values, such as negative or excessively high amounts. Let \( M(S') \) represent the outcome of setting these milestones. We expect that that \( M(S) \neq M(S') \) where \( M(S') \) should reject any milestone with invalid reward values, ensuring the contract enforces validity checks on milestone rewards to maintain reasonable and correct configurations.

\item \textbf{MR15: Consistency for Sequential Organizers' Donations} 
Consider a source test case \( S \) where organizers make their donations at spaced intervals, and let \( T(S) \) represent the state of the contract, transitioning to the "DONATION" state after all organizers have donated. Create a follow-up test case \( S' \) where organizers donate in rapid succession with minimal time gaps between each donation. Let \( T(S') \) represent the contract's state under these rapid donations. We expect that \( T(S') = T(S) \), confirming that the contract transitions to the "DONATION" state correctly, even when donations are made in quick succession.

\item \textbf{MR16: Consistency for Contract Closure with Pending Milestones}  
Consider a source test case \( S \) where an attempt is made to close the contract after all milestones have been either completed or refunded, and let \( C(S) \) represent the successful closure of the contract. Create a follow-up test case \( S' \) where an attempt is made to close the contract with milestones that are uncompleted or not refunded. Let \( C(S') \) represent the closure outcome in this follow-up case. We expect that \( C(S) \neq C(S') \) where \( C(S') \) should fail, ensuring the contract only allows closure when all milestones are resolved.

\item \textbf{MR17: Consistency for Repeated Withdrawals}

Consider a source test case \( S \) where each beneficiary withdraws their allocated funds once, and let \( W(S) \) represent the successful withdrawal status of each beneficiary. Create a follow-up test case \( S' \) where the same beneficiary attempts multiple withdrawals after the first successful one. Let \( W(S') \) represent the outcome of these repeated withdrawal attempts.
We expect that \( W(S)= W(S') \) where  \( W(S') \) accepts any withdrawal attempts by a beneficiary after the first successful withdrawal, allowing multiple withdrawals.

\end{itemize}

\section{Experimental Setup}
In this section we provide the details of the experimental setup,
especially the research questions to be answered and the system under test (SUT).
\subsection{Research Question}
\label{sec:research_question}
RQ1.  How effectively does MT identify faults in the SUT?

RQ2. Which MRs perform better in identifying faults in the SUT?

\subsection{Mutation Generation}
Mutation testing is used in our experiments to determine the quality and effectiveness of MRs to identify faults in the smart contract program. Mutants are generated for the program using the Vertigo mutation testing tool. The faulty versions of the program are created using mutation operators. Mutation operators apply changes to a statement in the program that creates a fault in the program. Table~\ref{tab:mutation_oprerator} describes the mutation operators used by Vertigo to generate mutants and generated 230 mutants combined for smart contract programs such as CrowdfundingCampaign.sol, CrowdfundingCampaignMilestoneSystem.sol and its associated libraries.

\begin{table}[h!]
\centering

\caption{Mutation Operators}
\label{tab:mutation_oprerator}
\begin{tabular}{|p{3cm}||p{5cm}|} 
\hline
\textbf{Mutation Operator} & \textbf{Description} \\
\hline
Condition Boundary & Replaces a conditional operation with its inclusive or exclusive counterpart \\
\hline
Condition Negation & Replaces a conditional operation with its inverse \\
\hline
Math Inversion & Replaces a math operator with its inverse \\
\hline
Increments Inversion & Replaces an increments statement with its inverse \\
\hline
Increments Mirror & Replaces an increments statement with its mirror \\
\hline
Modifier Removal & Removes a modifier application \\
\hline
\end{tabular}
\end{table}

\begin{table}[h!]
\centering
\caption{\# Mutants killed by each MR }
\label{tab:mutants_killed_mr}
\begin{tabular}{|p{1cm}|p{3cm}|p{3cm}|} 
\hline
\textbf{MR\#} & \textbf{\# Mutants Killed} & \textbf{Total Mutants for each MR} \\ \hline
1             & 69                         & 77                                 \\ \hline
2             & 69                         & 77                                 \\ \hline
3             & 3                          & 202                                \\ \hline
4             & 30                         & 202                                \\ \hline
5             & 20                         & 28                                 \\ \hline
6             & 27                         & 93                                 \\ \hline
7             & 25                         & 68                                 \\ \hline
8             & 0                          & 77                                 \\ \hline
9             & 38                         & 202                                \\ \hline
10            & 16                         & 24                                 \\ \hline
11            & 15                         & 51                                 \\ \hline
12            & 26                         & 202                                \\ \hline
13            & 38                         & 202                                \\ \hline
14            & 25                         & 100                                \\ \hline
15            & 20                         & 51                                 \\ \hline
16            & 19                         & 55                                 \\ \hline
17            & 22                         & 90                                 \\ \hline
\end{tabular}
\end{table}

\section{Results}
\begin{figure}[ht]
\centering
\includegraphics[width=0.5\textwidth]{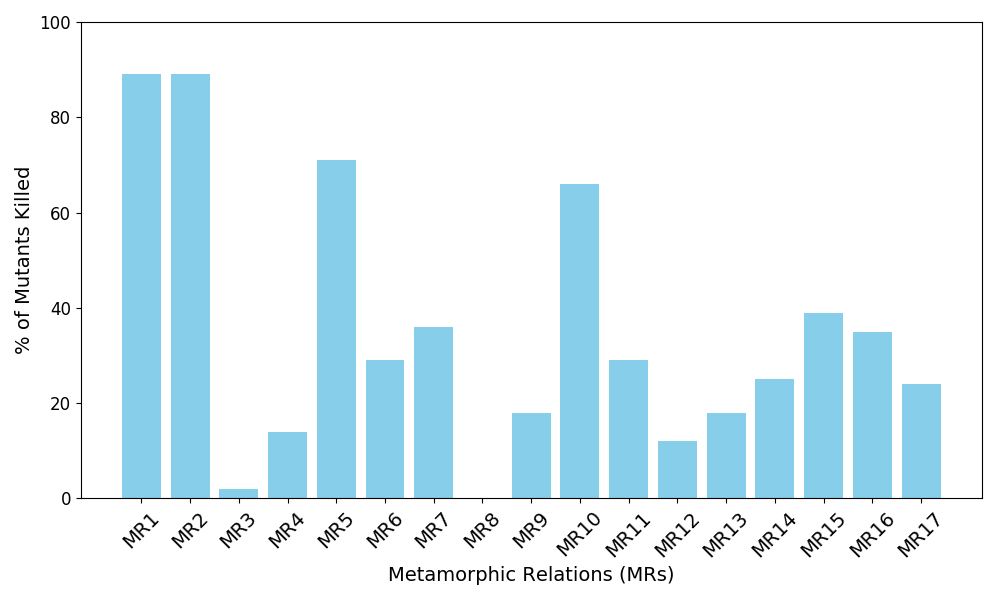}
\caption{Fault detection rate of MRs for Ethereum Crowdfunding Contract Program}
\label{fig:mutant_kill_rate_mrs}
\end{figure}

In this section, we discuss our findings for each of the research
questions described in section~\ref{sec:research_question}.
The results from Figure~\ref{fig:mutant_kill_rate_mrs} and Table~\ref{tab:mutants_killed_mr} illustrate the effectiveness of MRs in identifying faults in the smart contract program by measuring their mutant killing rates. For the result processing of each MR, only Killed and Alive mutants were considered. Mutants that resulted in execution errors during testing were excluded from the analysis to ensure the accuracy of the fault detection evaluation. 
The experimental results illustrate the effectiveness of MT in identifying faults in the Ethereum-based crowdfunding smart contract. The mutant killing rates of defined MRs indicate varying levels of success, emphasizing their role in addressing specific functionalities and fault scenarios within the smart contract.

Among the MRs, \textbf{MR1 (State Transition Consistency)} and \textbf{MR2 (Donations Consistency)} exhibit the highest mutant-killing rate of \textbf{90\%} accounting for a significant proportion of fault detection. \textbf{MR1} ensures the contract transitions through its lifecycle states correctly (\textit{STARTED} $\rightarrow$ \textit{DONATION} $\rightarrow$ \textit{ENDED}), while \textbf{MR2} validates that the total donation amount remains accurate regardless of the contribution method. Their effectiveness can be attributed to their alignment with critical and frequently exercised functionalities, making them robust in detecting a wide range of faults.

Moderate performance is observed in \textbf{MR9 (Minimum and Zero Donations)} and \textbf{MR13 (Overflow and Underflow Protection)}, each achieving a mutant killing rate of \textbf{ 16. 45\%}. \textbf{MR9} ensures donations meet minimum thresholds, while \textbf{MR13} verifies contract robustness against extreme transaction values, such as large or minimal amounts. These MRs address boundary conditions and edge cases, highlighting their importance in exposing specific but less frequent faults.

MRs targeting niche scenarios or stress conditions, such as \textbf{MR3 (Deployment Consistency with Duplicate Beneficiaries)} and \textbf{MR10 (Rapid State Transitions)}, show limited effectiveness, with mutant-killing rates of \textbf{1.30\%} and \textbf{6.93\%}, respectively. The low fault detection rate of \textbf{MR3} suggests that duplicate beneficiary scenarios are rare or well handled by the contract design. Similarly, \textbf{MR10}, which tests the behavior of the contract under rapid state transitions, reflects the limited exposure of faults in well-constructed systems.

\textbf{MR8 (Maximum Organizer or Beneficiary Limit)} stands out for its ineffectiveness, failing to detect any mutants. This result may be due to the contract's inherent robustness in enforcing participant limits or the MR's inability to target meaningful edge cases effectively.

The variability in MR performance highlights the importance of designing relations that align with areas prone to faults. High-performing MRs like \textbf{MR1} and \textbf{MR2} are indispensable to ensure the correctness of core functionalities, while moderately effective MRs contribute by addressing less frequent but critical edge cases. In contrast, low-performing MRs may require refinement to improve their relevance and impact.

These findings underscore the importance of a balanced testing strategy that prioritizes MRs targeting core operations while maintaining diversity to ensure comprehensive fault coverage. The study demonstrates the utility of MT in validating smart contracts, with results supporting continued refinement and expansion of MRs to improve testing efficacy.
\textbf{Answer to RQ1 and RQ2:} The results demonstrate that MT is effective in identifying faults in the smart contract program. The effectiveness of individual MRs varies significantly, with \textbf{MR1 (State Transition Consistency)} and \textbf{MR2 (Donations Consistency)} emerging as the most effective.

\section{Threats to Validity}

This section discusses potential threats to the validity of the findings in this study.

\subsection{Internal Validity}
One key threat to internal validity is the accuracy of the mutant generation process. The mutants used in this study may not fully represent real-world faults, which could lead to an overestimation or underestimation of the Metamorphic Relations' (MRs) effectiveness. Additionally, the implementation of MRs may introduce bias, as certain MRs might be designed to align closely with specific functionalities, inadvertently favoring those areas in fault detection.

\subsection{External Validity}
The generalizability of the results is limited by the specific context of the smart contract under test, which is an Ethereum-based crowdfunding application. Although MRs were effective in identifying faults within this context, their applicability to other types of smart contracts or blockchain platforms remains uncertain. More studies involving diverse smart contract designs are necessary to validate the broader utility of the proposed approach.

\section{Future Work}

Future work will focus on refining low-performing MRs to improve their relevance and fault-detection capabilities. This includes analyzing their limitations and redesigning them to better target specific fault scenarios or edge cases. Additionally, new MRs will be developed to address underexplored functionalities in smart contracts, ensuring comprehensive coverage and robustness in testing.

Automation is a critical next step to enhance efficiency. Machine learning or heuristic-based methods can be explored to automate the identification of MRs and the generation of source and follow-up test cases. This will streamline the testing process and reduce manual intervention, enabling faster and more scalable application of Metamorphic Testing.
 
The methodology can be extended to test a wider range of smart contracts, including those used in decentralized finance (DeFi), supply chain management, and healthcare. This expansion will help assess the generalizability and adaptability of the approach to various domains, ensuring its utility across diverse blockchain applications.

\section{Conclusion}

This study highlights the effectiveness of MT as a robust approach to identifying faults in smart contract programs. Using carefully designed MRs, MT addresses the test oracle problem inherent in smart contract testing, providing a reliable means of verifying contract behavior without requiring predefined expected outcomes. The results demonstrate that MRs targeting core functionalities, such as state transitions and donation consistency, exhibit the highest fault detection rates, underscoring their critical role in ensuring the correctness of smart contract operations. 

Moderately effective MRs contribute to fault detection by addressing boundary conditions and edge cases, while variability in effectiveness between different MRs highlights the need for a balanced and diverse testing strategy. The study emphasizes the importance of refining low-performing MRs and developing additional relations to enhance coverage across core and niche functionalities.

The findings provide a foundation for future work to expand the applicability of MT in smart contract testing, particularly through automation, real-world fault analysis, and integration with complementary testing techniques. By addressing these areas, MT can become an even more powerful tool to ensure the reliability, security, and robustness of smart contracts in blockchain systems. This research represents a significant step toward improving the verification before the implementation of smart contracts, reducing risks, and enhancing trust in blockchain-based applications.

\bibliographystyle{plain}
\bibliography{reference}


\end{document}